\documentclass[aps,reprint,amsmath, amssymb,groupedaddress,floatfix]{revtex4-1}
\usepackage{natbib}
\usepackage{graphicx}
\usepackage{bm}
\usepackage{color,soul}
\usepackage{hyperref}
\usepackage{float}
\usepackage{csquotes}
\usepackage{color,soul}
\usepackage{hyperref}
\hypersetup{colorlinks=true, urlcolor=blue, citecolor=blue}
\usepackage{longtable}
\usepackage{tabularx}
\newcommand{\cofor}{{\rm [NH$_2$NH$_3$]Co(HCOO)$_3$}}
\newcommand{\mnfor}{{\rm [NH$_2$NH$_3$]Mn(HCOO)$_3$}}

\begin{document}

\title{Coupling  of structure and  magnetism to spin splitting in hybrid organic-inorganic perovskites}

\author{Ravi Kashikar$^1$ }
\email{ravik@usf.edu}
\author{ D. DeTellem$^1$}
\author{P. S. Ghosh$^2$ }
\author{Yixuan Xu$^3$}
\author{S. Ma$^3$}
\author{S. Witanachchi$^1$}
\author{Manh-Huong Phan$^1$}
\author{ S. Lisenkov$^1$ }
\author{I. Ponomareva$^1$}
\email{iponomar@usf.edu}
\affiliation{1. Department of Physics, University of South Florida, Tampa, Florida 33620, USA}
\affiliation{2. Glass \& Advanced Materials Division, Bhabha Atomic Research Centre, Mumbai 400 085, India}
\affiliation{3. Department of Chemistry, University of North Texas, CHEM 305D, 1508 W Mulberry Street, Denton, Texas 76201, USA}

\begin{center}
    \date{\today}

\end{center}

\begin{abstract}
Hybrid organic-inorganic perovskites are famous for the diversity of their chemical compositions, phases and phase transitions, and associated physical properties. We use a combination of experimental and computational techniques to reveal strong coupling between structure, magnetism, and spin splitting  in a representative of the largest family of hybrid organic-inorganic perovskites: the formates. With the help of first-principles simulations, we find spin splitting in both conduction and valence bands of \cofor\, induced by spin-orbit interactions, which can reach up to 14~meV. Our magnetic measurements reveal that this material exhibits canted antiferromagnetism below 15.5 K. The direction of the associated antiferromagnetic order parameter is strongly coupled with the spin splitting already in the centrosymmetric phase, allowing for the creation and annihilation of spin splitting through the application of a magnetic field. Furthermore, the structural phase transition into experimentally observed polar Pna2$_1$ phase completely changes the aforementioned spin splitting and its coupling to magnetic degrees of freedom. This  reveals that in \cofor\, the structure and magnetism are strongly coupled to spin splitting in a way that allows for its manipulation through both magnetic and electric fields. As an example, for a given point inside the Brillouin zone of centrosymmetric Pnma phase of \cofor, spin splitting can be turned on/off by aligning the antiferromagnetic vector along certain crystallographic directions or through inducing a polar phase by the application of an electric field.  In addition, the spin textures in \cofor\, are fully coupled to these order parameter vectors. We believe that our findings offer an important step toward fundamental understanding and practical applications of materials with coupled properties.

\end{abstract}
\maketitle

\section{Introduction}

The ability to couple structural and magnetic degrees of freedom to materials properties is of tremendous scientific and technological importance. For example, coupling between structural  and electron spin degrees of freedom is promising for applications in spintronics\cite{AFM_Spin1,Jungwirth2016}. One of the ways to access spin degrees of freedom is through spin-orbit interaction, which breaks the Hamiltonian symmetry and splits spin-degenerate bands\cite{dresselhaus1955spin, even2013importance}. Spin-orbit interactions in solids give origin to the so-called Rashba and Dresselhaus effects, which originate from structural inversion asymmetry and bulk inversion asymmetry, respectively\cite{rashba1960properties,dresselhaus1955spin,Soumyanarayanan2016, Bihlmayer2022}. These effects give origin to spin splitting in the electronic bands and, consequently, the opportunity to access a particular spin channel and also lead to the emergence of spin texturing in either real or reciprocal space. Certain types of spin textures are known to enhance spin states' lifetimes and, therefore, are critical for applications in spintronics\cite{schliemann2017colloquium, tao2018persistent,  tao2021perspectives}. Structural degrees of freedom are typically associated with ionic positions and lattice vectors and can be manipulated through temperature and external fields\cite{Lines-Glass}. Ferroics usually offer great opportunities for structural manipulations as they have a tendency to undergo phase transitions and often couple to electric fields.  Magnetic degrees of freedom are usually associated with localized spins and may be either disordered or ordered.  Examples are paramagnetism and (anti) ferromagnetism, respectively. While ferromagnetism results in spin dependent electronic band structure, more subtle effect of spin splitting has been predicted from group theoretical analysis in antiferromagnetic (AFM) materials\cite{AFM_SS1,AFM_SS2,AFM_SS3,AFM_SS4,AFM_SS5,AFM_SS6,AFM_SS7,AFM_SS8}. Remarkably, some AFM materials can exhibit spin splitting even in the absence of spin-orbit interactions\cite{AFM_SS1}. It appears that materials in the ferroic, or more precisely multiferroic,  family may have significant  potential for coupling structure and magnetic ordering to spin splitting. 
This motivated us to look into the hybrid organic-inorganic perovskites class of materials, that share the chemical formula ABX$_3$, where A typically is an organic cation, while B is an inorganic cation, while X could be either organic or inorganic\cite{Stroppabook}. In particular, we will look into one of the largest families in this class: hybrid formate perovskites with chemical formula AB(HCOO)$_3$\cite{JCP_Formate}. What makes them attractive for our search of materials with coupled degrees of freedom, is that many of them exhibit AFM ordering and many undergo phase transitions into polar space groups above room temperature\cite{Multi-Formate}. For example, [NH$_2$NH$_3$][Co(HCOO)$_3$] and [CH$_3$NH$_2$NH$_2$][Mn(HCOO)$_3$] exhibit transitions into polar phases below 363 K and 309~K, respectively \cite{Expt_ferro, Mn-MOF}. Ferroelectricity in [NH$_2$NH$_3$][Co(HCOO)$_3$] was reported in Ref.\cite{PSGhosh_PRL}, while ferroelectric ordering in [CH$_3$NH$_2$NH$_2$][Mn(HCOO)$_3$] was suggested from pyroelectric measurements \cite{Mn-MOF}.   It is important to realize that in such materials magnetic ordering is mostly induced by the magnetic moments of the transition metal atoms on the B site, while ferroelectric polarization originates from the dipole moment of the A site, which may allow researchers to overcome the well-known limitations of inorganic materials for coexistence of magnetic and electric orderings \cite{Nicola}. Indeed, we can find multiferroics with desired features of AFM  in [(CH$_3$)$_2$NH$_2$]M(HCOO)$_3$ (M=Mn, Fe, Co, Ni ) and NH$_4$M(HCOO)$_3$ (M=Mn, Co) family \cite{Multi-Formate}. Among them, one promising candidate for coupling between multiple degrees of freedom to spin splitting  is \cofor. It exhibits a high-temperature Pnma phase and undergoes a phase transition to the polar Pna2$_1$ phase at 363 K \cite{Co_formate_expt}. As a result, it is predicted to be ferroelectric at room temperature with a remnant polarization of 2.6  $\mu$C/cm$^2$ \cite{PSGhosh_PRL}. The structural phase transition in this compound is driven by the hydrogen bond stabilization \cite{PSGhosh_PRL}. \cofor\, also exhibits competing, possibly antipolar, P2$_1$2$_1$2$_1$ phase, which  show spin-canted AFM long-range-ordering, with Néel temperatures of 13.9 K \cite{Expt_ferro}. Therefore, it may exhibit the desired coupling of structure and  magnetism to spin splitting.

Previously spin splitting coupled to the direction of polarization and AFM order parameters, tunable by both magnetic ordering and polarization direction,  has been predicted in hybrid organic-inorganic perovskite  TMCM-MnCl$_3$\cite{Lou2020}.  Recently,  Yananose et al., have examined C(NH$_2$)$_3$]M(HCOO)$_3$ (M = Cr, Cu) and predicted the persistent and irregular spin texture \cite{acs_Stroppa}. Among nonmagnetic hybrid perovskites TMCM-CdCl$_3$ and MPSnBr$_3$ exhibit persistent spin texture in the valence band and conduction band respectively\cite{TMCM-PRB,Ravi_PRM}. However, what is presently missing is the understanding of the interplay between structural and magnetic degrees of freedom in the establishment of spin splitting in hybrid organic-inorganic perovskites.

 The goal of this study is to utilize a combination of computational and experimental tools to: (i) predict the existence of spin splitting in hybrid formate perovskites, which is strongly coupled to both structural and magnetic degrees of freedom; (ii) reveal the possibility to create, annihilate and manipulate spin splitting through application of electric and magnetic fields; (iii)  rationalize our findings through decoupling contributions from different degrees of freedom; (iv) propose practical ways to achieve spin splitting manipulation in such materials.

\section{ Experimental and Computational Details}

\cofor\ was synthesized using the mild solution method. The solvent diffusion was carried out using Co(No$_3$)2.6H$_2$O (1.0 mmol, 0.291 g) in 5.0 mL methanol solution. Later, it was layered onto 5.0 mL methanol with 95\% (78 mmol, 3.2 mL) formic acid and 98\% (6.2 mmol, 0.6 mL) hydrazine monohydrate. The magenta-colored rectangular crystals are obtained after 24 hours which are separated from the bulk phase and washed with reagent grade ethanol after the mother solvent was
removed. The perovskite structures are confirmed with single-crystal X-ray diffraction measurements.  Magnetic measurements were carried out in a Quantum Design Dynacool Physical Property Measurement System (PPMS) using a vibrating sample magnetometer (VSM) with the magnetic field aligned parallel to the a-b plane. Magnetization versus applied field M(H) was measured at 10 K, with an applied field of 9 T. Magnetization versus temperature,  $M(T)$, was measured between 5 K and 100 K, using the zero-field-cooled warming protocol and under 0.1 T applied field.

For computations we use spin-polarized density functional theory as implemented in the Vienna Abinitio Simulation Package (VASP) with a Perdew-Burke-Ernzerhof (PBE) exchange-correlation functional \cite{VASP_1, VASP2, PBE}.  The energy cutoff is set to 700 eV for the plane-wave basis, and the Brillouin zone integration is carried out using Gamma-centered k-mesh of 5 $\times$ 5 $\times$ 3.  We use U$_{eff}$ = 2.0 eV correction for Co-d states as proposed by Dudarev et al. \cite{Duradev}. The zero-damping DFT-D3 approach, proposed by Grimme et al., is employed to incorporate the van der Waals interactions\cite{Grimme}. 
Structural relaxations are performed using a conjugate gradient algorithm until the forces on each atom are less than 1 meV/\AA, while stresses are less than 0.001 GPa.    The electronic structure calculations are performed with both spin-orbit coupling  (SOC) on and off. The electric polarization is computed  using the modern theory of polarization developed by King-Smith and
Vanderbilt\cite{king_smith}. 

  \begin{figure*}
\centering
\includegraphics[width=1.0\textwidth]{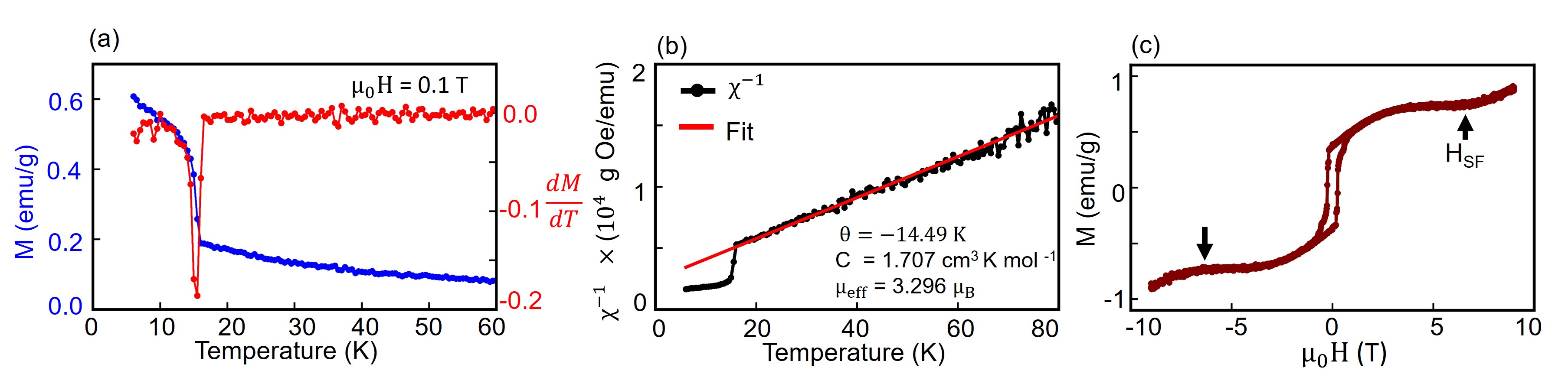}
\caption{ (a) Temperature dependence of the magnetization (blue) and its derivative dM/dT (red); (b) Inverse susceptibility as a function of temperature $\chi^{-1}$(T) and a fit to the Curie-Weiss law (red line); (c) Magnetic field dependence of magnetization curve taken at 10 K, showing a ferromagnetic hysteresis at low fields and a spin-flop transition around 6.5 T.    }
\label{fig1}
\end{figure*}

\section{Results and Discussions}

Although AFM ordering has been reported for \cofor\ in P2$_1$2$_1$2$_1$ phase \cite{Expt_ferro}, no reports are available for the polar Pna2$_1$ phase. Our magnetic data analysis for the Pna2$_1$ phase are given in Fig.~\ref{fig1}  which suggest canted AFM ordering below  15.5 K.  Figure \ref{fig1}(a) shows the temperature dependence of magnetization along with its derivative,  $dM/dT$, revealing a magnetic phase transition at 15.5 K. Inverse susceptibility $\chi^{-1}$ was calculated from this data $H/M(T)$ and fit with the Curie-Weiss law, $ \chi = C/(T - \theta)$ (Fig. \ref{fig1}(b)), where a Curie-Weiss temperature of $\theta$ = -14.49 K is obtained. It is clear that at high temperatures in the paramagnetic regime, the $\chi^{-1}$(T) data obeys the Curie-Weiss law. The negative value of $\theta$ confirms
AFM couplings between adjacent Co$^{2+}$ ions and that the phase transition at T$_N$ $\sim$ 15.5 K is AFM in nature. From the Curie constant obtained through the fit (C = 1.707 cm$^3$ K mol$^{-1}$), an effective magnetic moment of $\mu_{eff}\sim$3.296 $\mu$B/Co is obtained, using the relation $C = N_A \mu_B^2 \mu_{eff}^2/3k_B$. It is also worth noting from Fig. \ref{fig1}(b) that there is a sudden decline in $\chi^{-1}$ at $\sim$15 K, signaling the characteristic of a spin-canted AFM system \cite{M1, inorganics11110444, M2}. To confirm the occurrence of spin canting in \cofor\, the magnetic field dependence of magnetization M(H) was measured at temperatures below the T$_N$. Fig. \ref{fig1}(c) displays such a M(H) loop taken at 10 K, confirming that there is a small canted AFM moment at low fields of $\sim$0.4 emu/g, with a coercive field of 0.25 T. When the applied magnetic field exceeds H$_{SF}$ $\sim$ 6.5 T, there is an increase in the magnetization characteristic of a spin-flop transition (Fig. \ref{fig1}(c)) \cite{M1}. This provides further evidence for the spin-canted AFM order in \cofor\ \cite{Expt_ferro}. The spin-canted AFM behavior was also observed in Cu$_3$(TeO$_4$)(SO$_4$).H$_2$O \cite{M1} and \{Co(N$_3$)(bpmb)(H$_2$O)$_2$.H$_2$O\}$_n$ \cite{inorganics11110444}. Thus, at low-temperature \cofor\ in Pna2$_1$ phase is a multiferroic as it exhibits the coexistence of electric (spontaneous polarization) and magnetic orderings. As a result, it could potentially exhibit spin splitting due to AFM ordering and noncentrosymmetric nature.

\begin{figure*}
\centering
\includegraphics[width=0.8\textwidth]{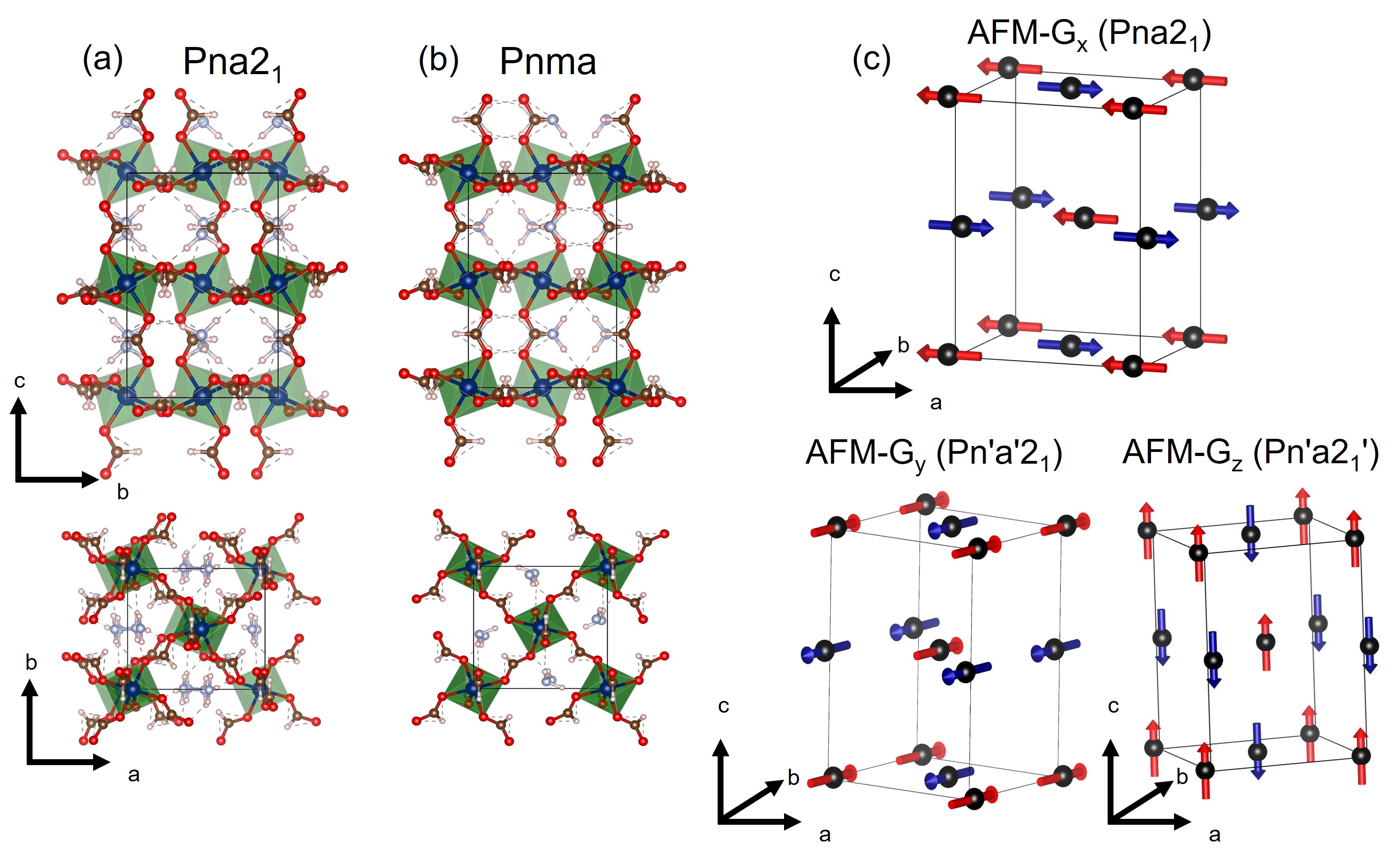}
\caption{  Crystal structure of \cofor\ in Pna2$_1$ and Pnma phases in different cross sections (a)-(b).  Magnetic moment arrangements in G-AFM ordering and corresponding magnetic space group (c).    }
\label{fig2}
\end{figure*}

  To find out whether these two orderings are indeed capable of inducing spin splitting in this material, we turn to simulations.  To decouple the contributions from magnetic and electric orderings, both centrosymmetric and noncentrosymmetric structures are required. To obtain the noncentrosymmetric phase, we carried out the structural relaxation of the Pna2$_1$ phase reported in
Ref. \cite{PSGhosh_PRL}.  The structure is visualized in Fig.~\ref{fig2}(a).  The ground state was shown to be G-type AFM in Ref. \cite{PSGhosh_PRL}. Therefore, we use G-type AFM arrangements of magnetic moments on Co atoms but simulate three different orientations of the magnetic moments:   along the -x, -y, and -z directions, as shown in Fig.~\ref{fig2}(c). Note that the axes of our Cartesian coordinate system are aligned along the crystallographic a, b, and c axis. It should be noted that we do not find canted magnetic moment in computations, which could be both due to our computational resolution and the size of the supercell.  The centrosymmetric structure for \cofor\, is Pnma which is also a high-temperature phase reported in Ref.\cite{PSGhosh_PRL}. However, it is a disordered phase with partial occupancies and, therefore, cannot be used in simulations. To construct a prototype of such a phase that does not have partial occupancy but still has experimental relevance, we create a polarization reversal path for the Pna2$_1$ phase following Ref.\cite{Maggie_inverse}. The structure that corresponds to zero polarization is the centrosymmetric Pnma phase. This structure is next subjected to full structural relaxation, which preserved the space group. The structure is visualized in Fig.~\ref{fig2}(b).

Experimentally, such a phase could occur during polarization reversal at the fields close to the coercive field. At zero Kelvin, the Pnma structure is 280 meV/f.u. higher in energy than the ground state. The structural relaxation has been carried out for different types of magnetic orderings: ferromagnetic,   A-type, C-type, and G-type AFM orderings, and it was found that G-type AFM ordering is the lowest in energy. Furthermore, we considered G-type AFM orderings with different directions of magnetic moments, resulting in the alignment of the AFM vector along x, y, or z-directions as shown in Fig.~\ref{fig2}(c).    We found that AFM-G$_x$ (Pnma)  has the lowest energy
 followed by AFM-G$_z$ (Pn$^\prime$ma$^\prime$) and AFM-G$_y$ (Pn$^\prime$m$^\prime$a),  with 0.002 and 0.02 meV/f.u. higher energy, respectively. The structures are provided in Ref.\cite{ourgithub}

\begin{figure*}
\centering
\includegraphics[width=1.\textwidth]{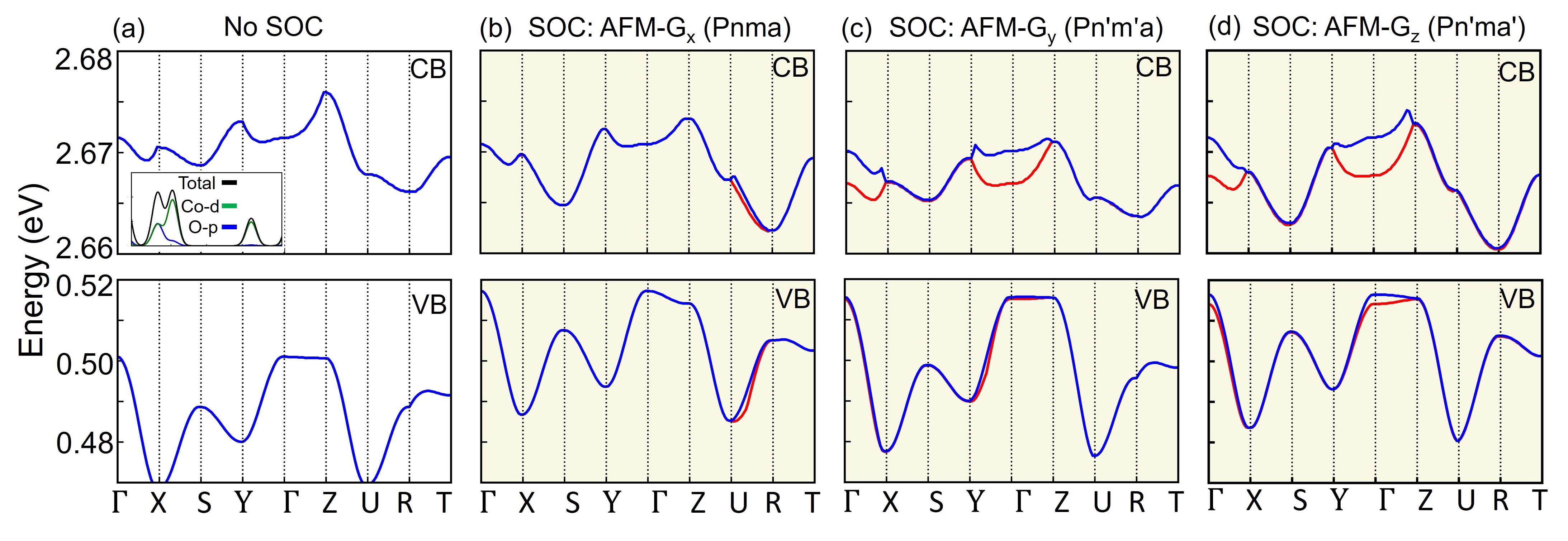}
\caption{Band structure of \cofor~in Pnma phase without (a)  and with (b)-(d) SOC.  }
\label{fig3}
\end{figure*}

The magnetic space groups Pn$^\prime$m$^\prime$a,  Pn$^\prime$ma$^\prime$, and Pnma do not possess $\Theta I$ symmetry and, therefore, may exhibit SOC-induced spin splitting \cite{AFM_SS1}. Here, $\Theta$ and $I$ are time reversal and spatial inversion symmetries.  Moreover, these magnetic space groups belong to types III, III, and I, respectively, which allows for AFM-induced spin splitting, as proposed in Ref. \cite{AFM_SS1}.   Following the classification of Ref. \cite{AFM_SS4} they belong to spin split type-4A and allow for spin splitting with or without SOC, based on the symmetry considerations. To explore that, we carry out DFT computations of electronic structure with SOC turned off and on. Of course, when SOC is turned off, we use spin-polarized computations. Figure \ref{fig3} presents our data for both cases. We find no spin splitting in the absence of SOC. The partial density of states shown in the inset to Fig.~\ref{fig3}(a)  infers that bands near the Fermi level are from Co-d states. In addition, the O-p states are populated at the valence band region. Among the Co-d states, d$_{xy}$ and d$_{z^2}$ have double occupancy, and hence, they peak in the valence band spectrum. The remaining d orbitals have single occupancy, thus offering a magnetic moment of 2.68 $\mu_B$  at each Co site. 

\begin{figure}[b]
\centering
\includegraphics[width=0.5\textwidth]{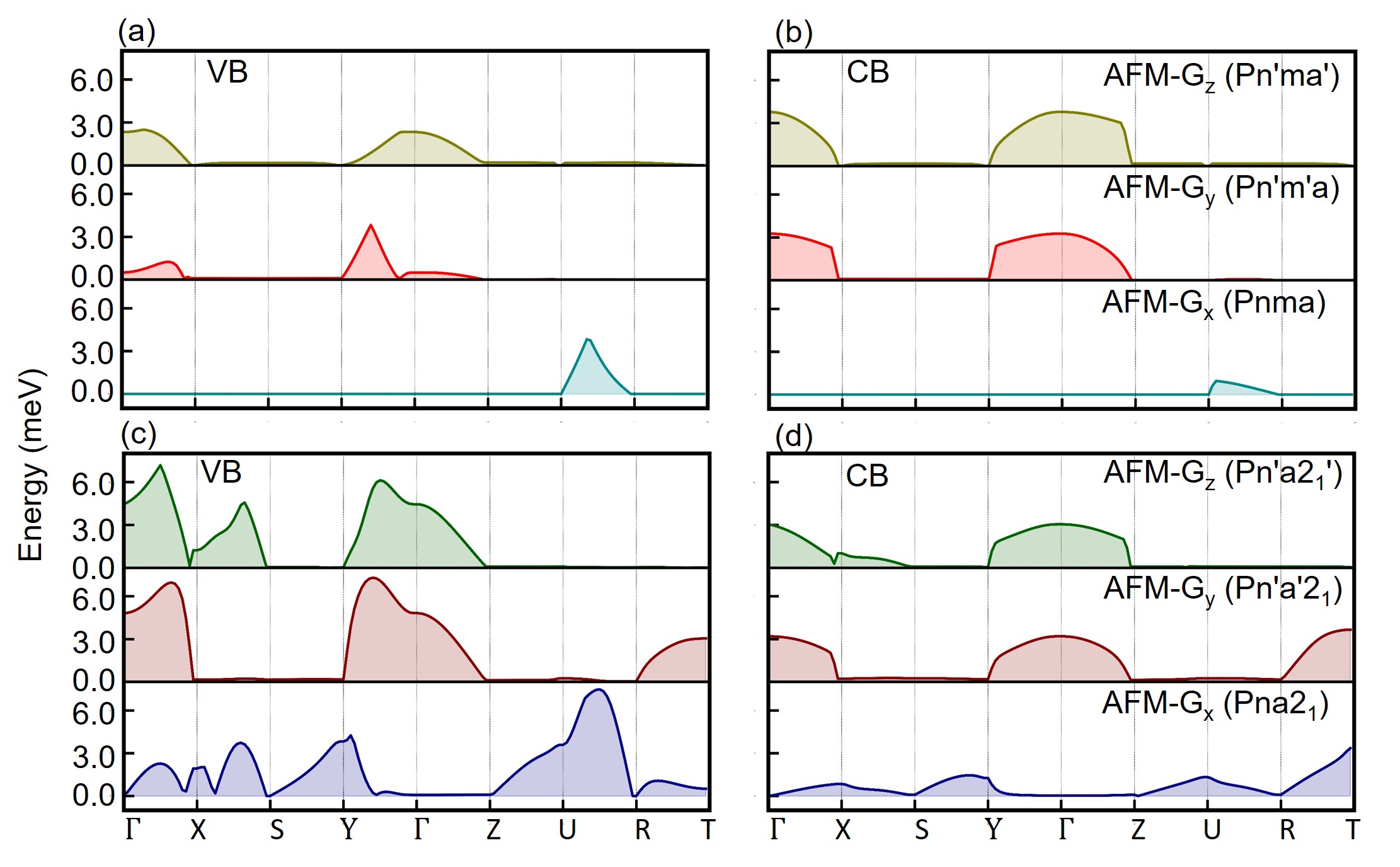}
\caption{  Spin splitting between the top two VBs (a) and (c) and bottom two CBs (b) and (d) along high symmetry k-path in different phases of \cofor. }
\label{fig4}
\end{figure}

Turning  SOC on results in spin splitting (see Fig.~\ref{fig3}(b)-(d)), which shows a strong dependence on the direction of the AFM order parameter. To quantify such dependence we compute spin splitting along the different directions of the Brillouin zone and present it in Fig.~\ref{fig4}.

Figure~\ref{fig4} confirms spin splitting dependence on the direction of AFM order parameter throughout the Brillouin zone. For example, the spin splitting along Y-$\Gamma$-Z direction in CB can be turned on or off by switching the AFM vector between the y or the z  direction. Similar trends are observed in VB.  Experimentally, the change of the AFM vector direction can be achieved through the spin-flop mechanism, which opens the possibility to create/erase or tune spin splitting by the application of a magnetic field. For instance, the field at which the spin-flop transition occurs is ~6.5 T for T = 10 K (Fig. \ref{fig1}(c)). We note that the magnitude of spin splitting also shows a strong dependence on the direction of the AFM vector. Thus, our simulations on the centrosymmetric phase prototype for AFM \cofor\, revealed: (i) the presence of SOC-induced spin splitting up to 4~meV in both CB and VB; (ii)  strong dependence of spin splitting magnitude on the direction of AFM vector, which allows for both creation and erasing of spin splitting, as well as its manipulations by the magnetic field.
\begin{figure*}
\centering
\includegraphics[width=1\textwidth]{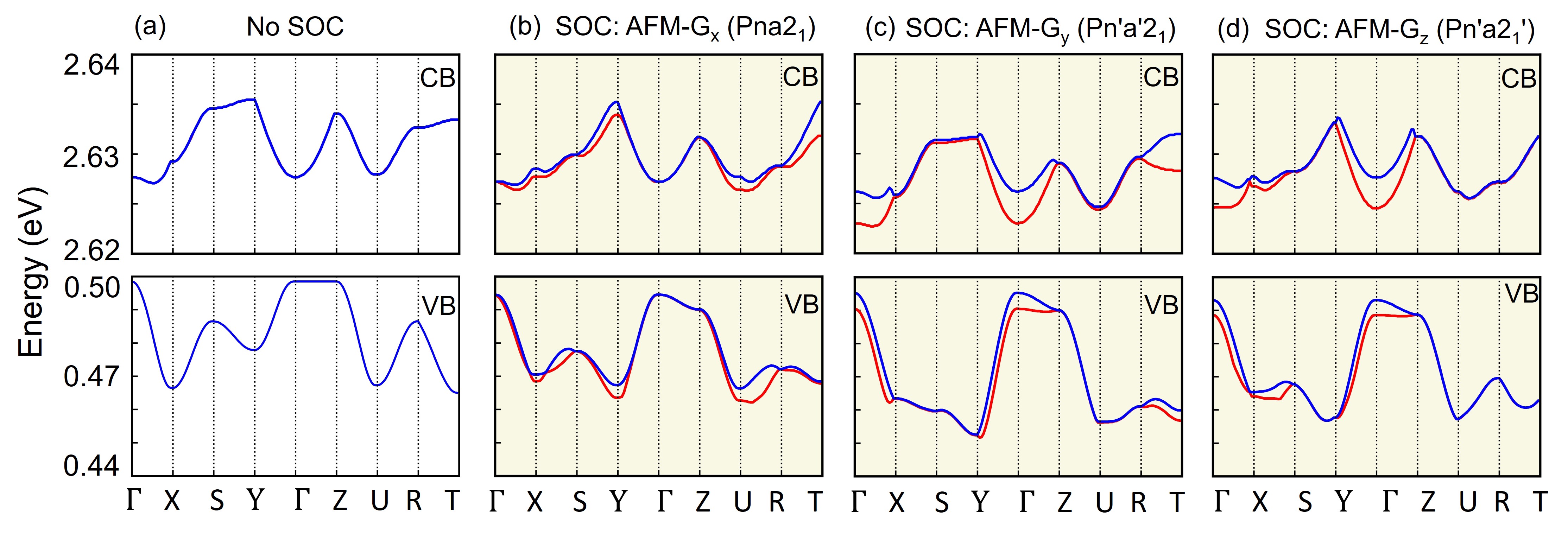}
\caption{  Band structure of Pna2$_1$ \cofor~ without SOC (a), and with SOC in (b)-(d) for x, y, and z spin orientations of AFM-G magnetic ordering.  }
\label{fig5}
\end{figure*}
Next, we turn to an investigation of the role that spatial asymmetry plays in spin splitting and the aforementioned findings.  Removal of the mirror plane perpendicular to the c-axis results in the ground state structure of \cofor\, with the Pna2$_1$ space group. Introducing G-type AFM ordering with different directions of $\mathbf G$ vector followed by full structural relaxation resulted in the structures with magnetic space groups Pna2$_1$, Pn$'$a$'$2$_1$, and Pn$'$a2$_1'$  for AFM-G$_x$, AFM-G$_y$, and AFM-G$_z$, respectively.  Energetically, AFM-G$_x$ has the lowest energy, followed by  AFM-G$_z$ (0.18 meV/f.u.) and  AFM-G$_y$ (0.47 meV/f.u.).  All these structures belong to the spin splitting type-4B as per classification of Ref.\cite{AFM_SS4},  and by symmetry, allow for spin splitting with and without SOC. Figure \ref{fig5} presents the electronic structure calculations for both of these cases.

Again, we find that just AFM on its own is not capable of inducing spin splitting in this material. However, the introduction of SOC results in such spin splitting for all directions of the AFM vector considered here. Moreover, just like before, we find that both magnitudes of spin splitting in its location in the Brillouin zone exhibit strong dependence on the direction of AFM vector. This is quantified in Fig.~\ref{fig4}. The magnitude of spin splitting is up to 7~meV in CB and up to 4~meV in VB, which offers enhancement with respect to centrosymmetric phases. We find a direct band gap of 2.1 eV, which is likely to underestimate the true band gap, as known for DFT. Remarkably, the bands are spin-splitted at VBM and CBM (located at $\Gamma$ point) despite being in time-reversal invariant momenta point. This is the consequence of the AFM ordering, discussed in Ref.\cite{AFM_SS1, AFM_SS4}. It is also a rare, but technologically significant property.
Comparison between  our data for centrosymmetric and noncentrosymmetric phases with different magnetic order parameter directions (see Fig.~\ref{fig4}) reveals that \cofor\, exhibits strong coupling between multiple degrees of freedom: structural, magnetic, and spin splitting.  For example,  in $\Gamma$ point of VB spin splitting is zero for AFM-G$_x$. It can be turned on by aligning the AFM vector along the y-axis, via a spin-flop mechanism induced by a magnetic field. Inducing the centrosymmetric phase through electric field or temperature manipulations will result in near disappearance of spin splitting. Figure ~\ref{fig4} also reveals greater overlap in the spin splitting peaks between space groups containing time-reversed symmetry elements as compared to those that don't, suggesting strong contributions to spin splitting from AFM.

\begin{figure*}
\centering
\includegraphics[width=0.8\textwidth]{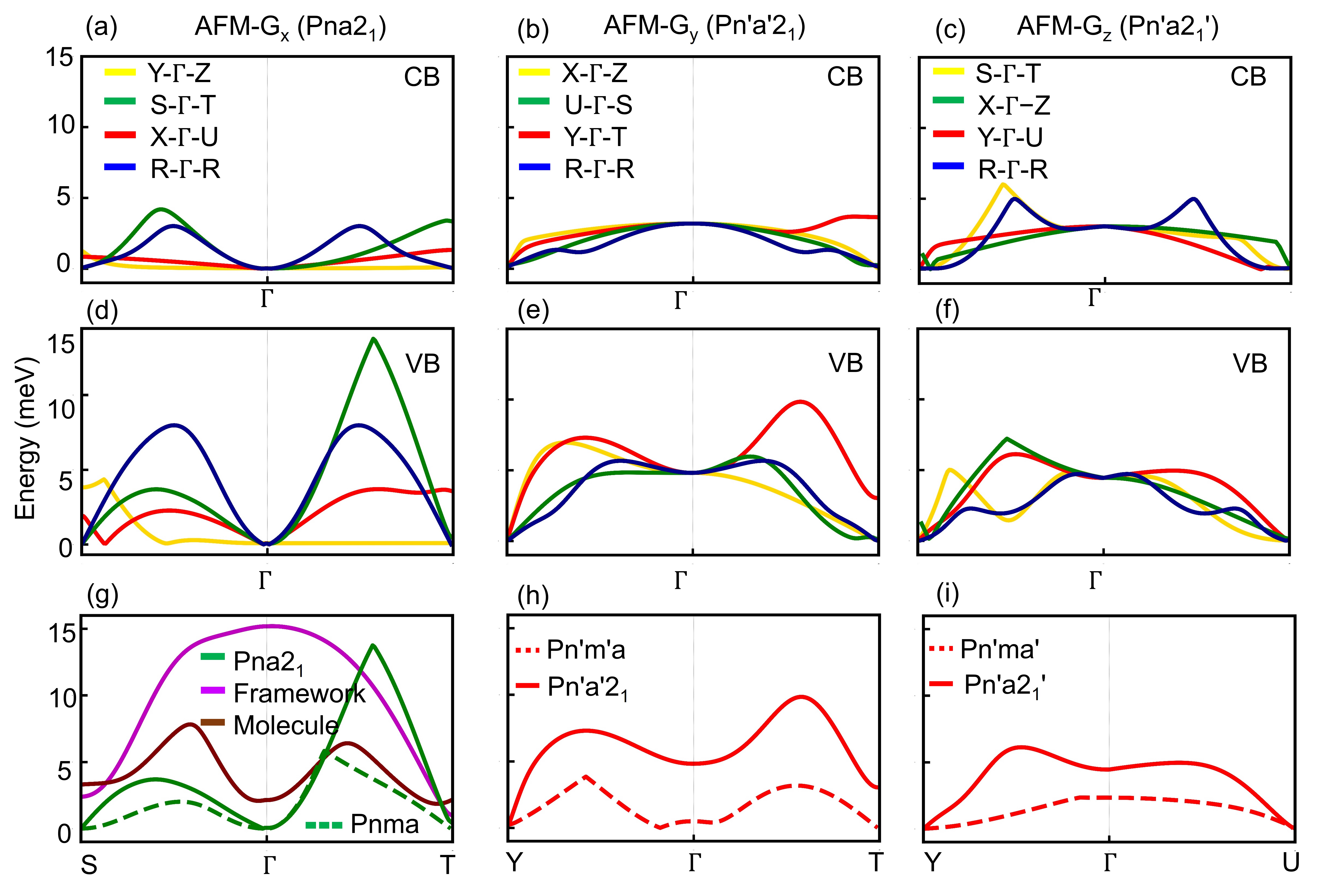}
\caption{ Spin splitting along various k-paths in the Brillouin of the polar phases of \cofor\, for the lowermost conduction bands (a)-(c) and uppermost valence bands (d)-(f) for x, y, and z orientations for the AFM order parameter.  Comparison of uppermost VB spin splitting for centrosymmetric and polar phases (g)-(i).   }
\label{fig6}
\end{figure*}

To find out whether even larger values of spin splitting can be achieved in the vicinity of VBM and CBM  as well as to assess anisotropy of the spin splitting, we explore several directions in the vicinity of CBM and VBM. Figure \ref{fig6} gives spin splitting along the directions investigated. The largest value can reach 14 meV in VB and occurs between $\Gamma$-T. To estimate the contribution to spin splitting from antiferromagnetism, we overlap the data for centrosymmetric and noncentrosymmetric structures in Figs. \ref{fig6}(g)-(i). For this analysis, the path in the Brillouin zone associated with the largest spin splitting is chosen. From the comparison, we conclude that AFM makes a significant contribution to spin splitting in all cases. Further analysis can be carried out to decouple structural contribution to spin splitting into contributions from the [NH$_2$NH$_3$]$^{-1}$ molecule and the Co(HCOO)$_3$ framework. To compute the contribution due to the molecule, we keep the framework in its Pnma structure while molecules are in their Pna2$_1$ configuration.  To compute contributions due to the framework, we keep molecules in their Pnma positions while the framework is in Pna2$_1$ structure. The decomposition is shown in Fig.~\ref{fig6}(g). We can see that the largest contribution to the spin splitting comes from the framework. For a tentative explanation of such a finding, we decompose the polarization of the Pna2$_1$ phase into contributions from the molecule and the framework. The framework contributes 73\% to the total polarization, while the molecule contributes 27\%. Therefore, it may be that structural distortions of the framework make the largest contribution to the local electric field, responsible for the spin-orbit interaction and associated spin splitting.

\begin{figure*}
\centering
\includegraphics[width=0.8\textwidth]{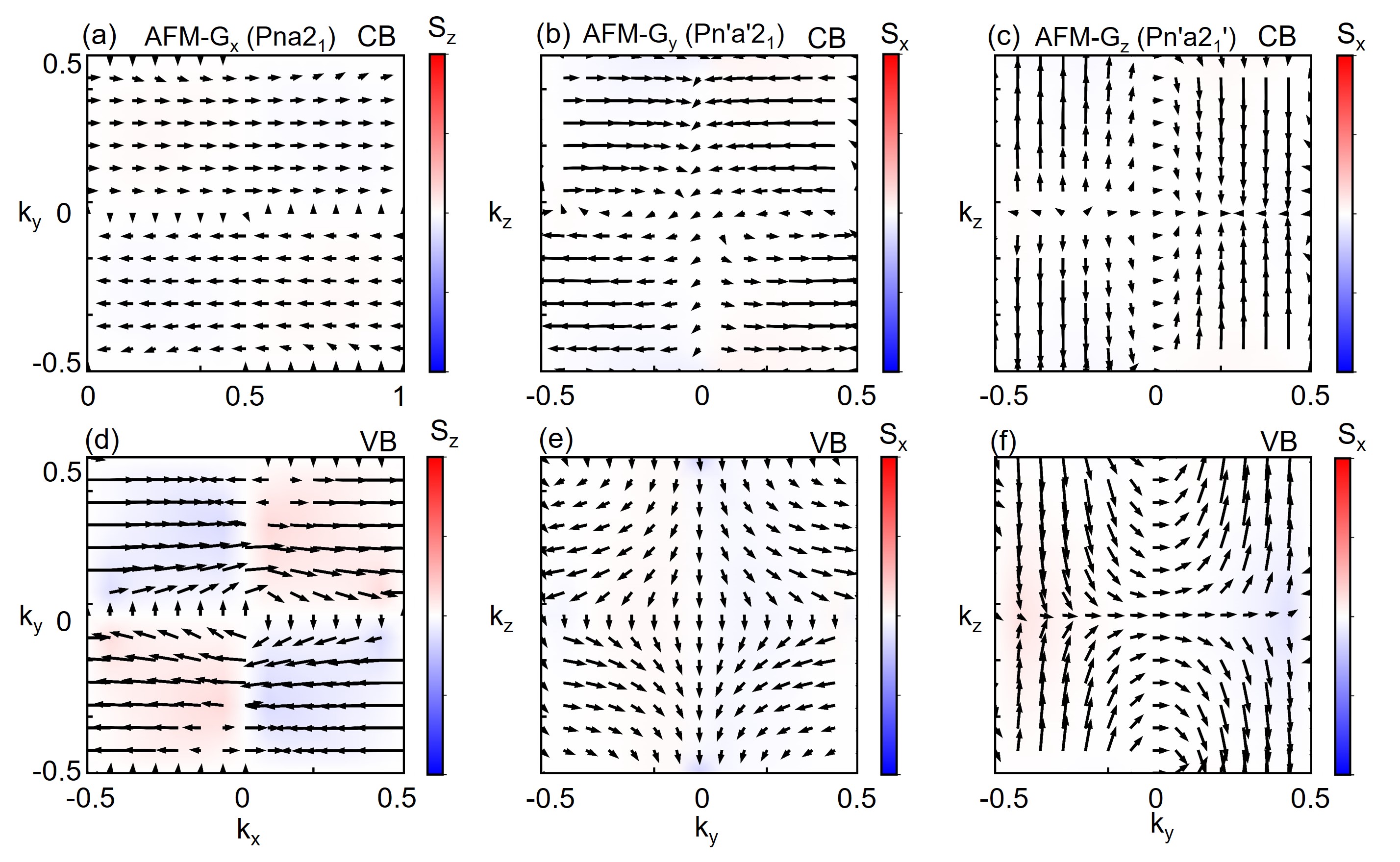}
\caption{ Spin textures around the CBM (a)-(c) and VBM (d)-(f) in various k-planes for the polar phases of \cofor. }
\label{fig7}
\end{figure*}

Figure \ref{fig7} shows spin textures in CB and VB for Pna2$_1$ type structures. We find that the type of spin texture is fully coupled to the direction of the AFM vector. Interestingly, for the Pna2$_1$ structure with AFM-G$_x$ orientation, the spin textures are mostly persistent in a large portion of the Brillouin zone which is highly desirable for the application of spintronics. Switching the direction of the AFM order parameter changes the type of spin textures and, therefore, will cause relaxation.

In order to find out whether the discussed findings are specific to \cofor\, or likely to occur in other materials from the same family we have repeated calculations for  \mnfor. This material exhibits a canted AFM phase below 7.9 K and undergoes a polar (Pna2$_1$) to nonpolar (Pnma) phase transition at 355 K\cite{Expt_ferro}, similar to the case of \cofor. All computational approaches remained the same except that we used  U$_{eff}$=4.0 eV for Mn. The structures are provided in Ref. \cite{ourgithub}. The simulated electronic structure for Pnma and Pna2$_1$ phases are shown in Fig. S1 and S2 of the Supplementary Material for -x, -y and -z orientation of the AFM order parameter. In both cases, we observe spin splitting in the conduction and valence bands. The magnitude of the spin splitting is 4 meV, near $\Gamma$ for VB for both phases. The magnitude of the spin splitting varies with magnetic moment directions. The dependence of spin splitting in  VB and CB on the direction of the AFM vector of both phases is shown in Fig. S3. The spin texture around the VBM for \mnfor\ is persistent in nature (see Fig. S4 of the supplementary material). Thus, \mnfor\, also exhibits the main features established for \cofor.  These include strong coupling between structural, magnetic, and spin degrees of freedom; the presence of spin splitting even in the centrosymmetric phase of the material; and persistent spin textures.

 \section{Summary}

 In summary, we have synthesized \cofor\, in the polar Pna2$_1$ space group and found it to be a canted antiferromagnet below 15.5 K. DFT computations have then been utilized to predict spin splitting in this compound. Interestingly, SOC-induced spin splitting exists already in the prototype of the centrosymmetric phase of the material and even in the time-reversal invariant momenta points of the Brillouin zone. These findings originate from the antiferromagnetic ordering. The spin splitting couples strongly to the direction of the AFM vector and can be turned on or off through AFM vector reorientation, which opens a way to spin splitting tunability by the application of a magnetic field, for example, via a spin-flop mechanism. 
 Phase transition from the centrosymmetric to the polar phase of the material results in a dramatic change in the spin splitting landscape in the Brillouin zone, thus offering additional tunability by the electric field or temperature. For example, a macroscopically centrosymmetric phase can be achieved at low temperatures during electric-field-induced polarization reversal.  Just like in the prototypical centrosymmetric phase of the material, the polar phase exhibits a strong dependence of the spin splitting on the direction of the AFM vector. Thus, our work reveals that the low-temperature polar phase of \cofor\, (and \mnfor) is magnetically ordered and exhibits strong coupling between structure, magnetic ordering, spin splitting and spin textures. These findings enrich our fundamental understanding of property manipulation and materials functionality and are likely to stimulate further research.  From a practical viewpoint, both structure and magnetic ordering can be manipulated by external fields, promising novel applications in spintronics.

\section*{Acknowledgment}
This work was supported by the National Science Foundation under Grant No. EPMD-2029800.

\bibliography{paper}

\newpage
\onecolumngrid

\begin{center}
   \textbf{\Large Supplementary Material}
\end{center}

 \vspace{1cm}

 \renewcommand{\thefigure}{S\arabic{figure}}

\setcounter{figure}{0}
 
\hspace{3cm}
\begin{figure}[h]
\centering
\includegraphics[width=1.\textwidth]{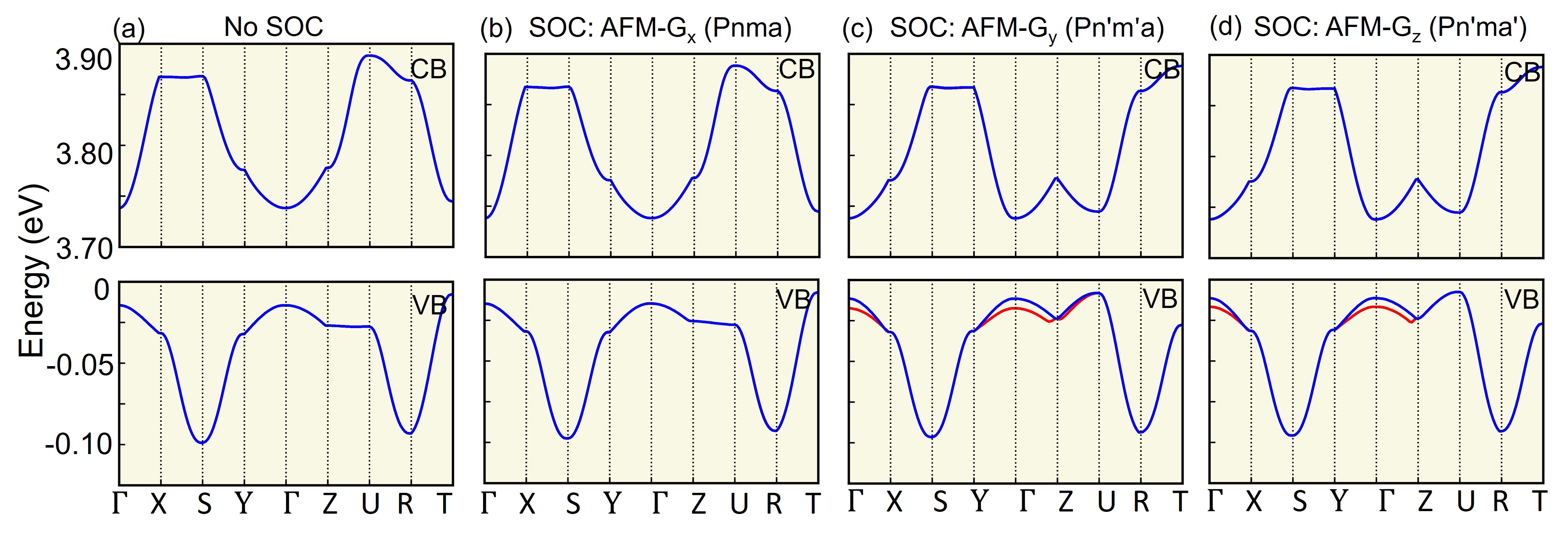}
\caption{Band structure of \mnfor~in Pnma phase with SOC.  }
\label{figS1}
\end{figure}

\begin{figure}[h]
\centering
\includegraphics[width=1.\textwidth]{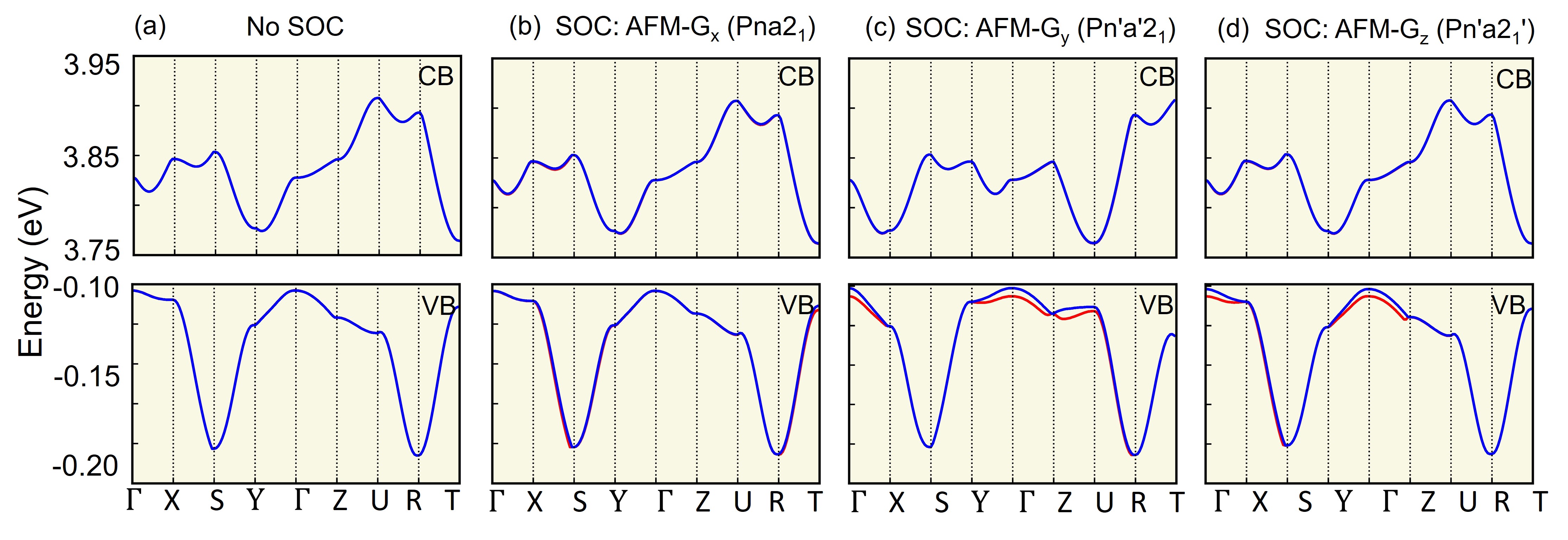}
\caption{Band structure of \mnfor~in Pna2$_1$ phase with SOC. }
\label{figS2}
\end{figure}

\begin{figure}[h]
\centering
\includegraphics[width=0.8\textwidth]{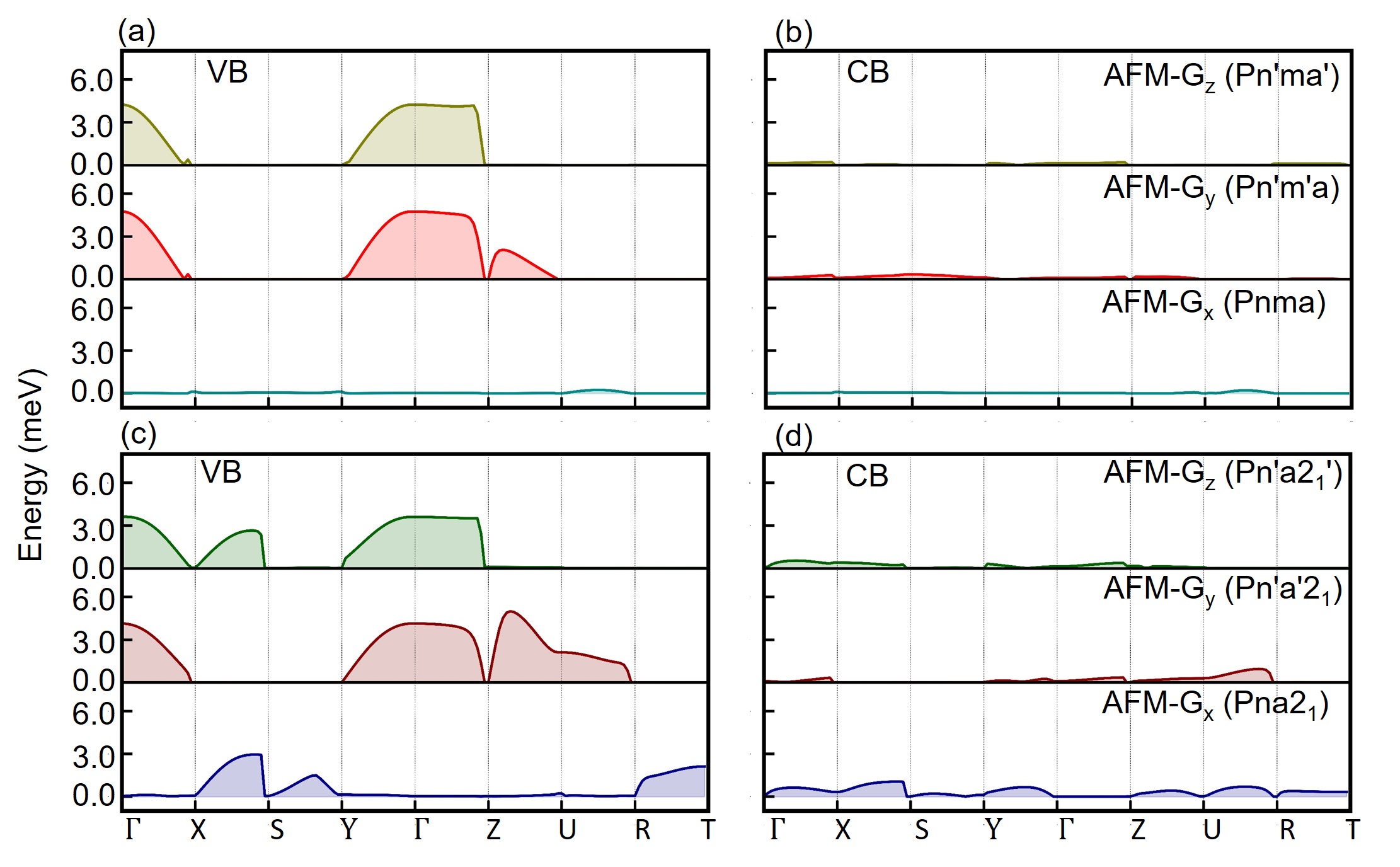}
\caption{Spin-splitting between the top two valence bands and bottom two conduction bands along high symmetry k-path in different phases of \mnfor.  }
\label{figS3}
\end{figure}

\begin{figure}[h]
\centering
\includegraphics[width=1.\textwidth]{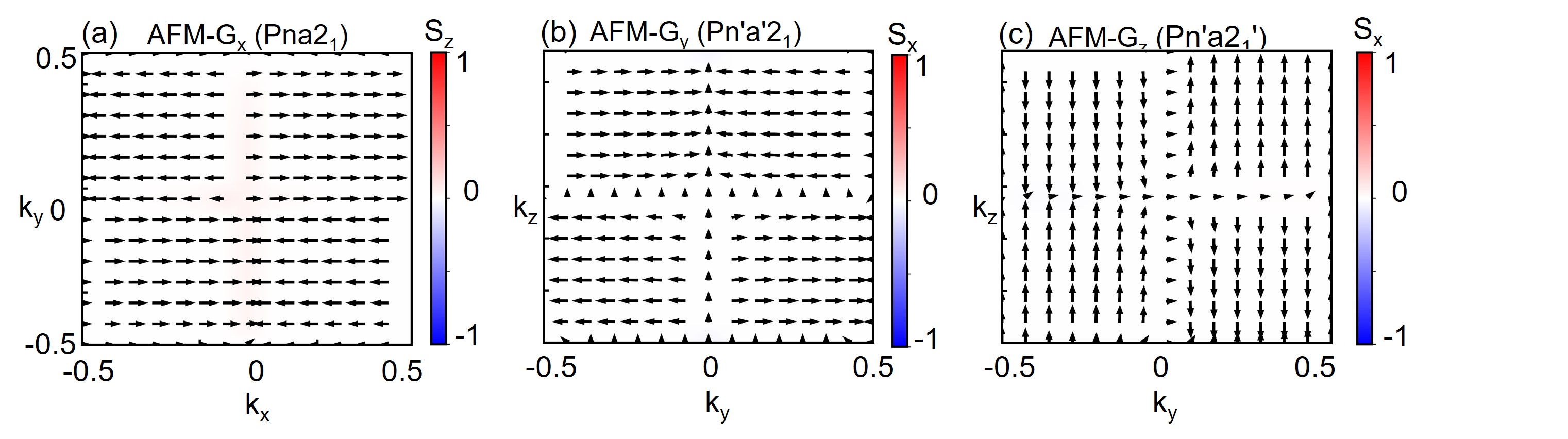}
\caption{ Spin textures of \mnfor around the valence band maximum in various k-planes. }
\label{figS4}
\end{figure}

\end{document}